\documentclass[iop]{emulateapj}
\usepackage{apjfonts}
\usepackage{natbib}
\def \msun{\hbox{$\rm{M}_\odot$}}

\newcommand{\bma}[1]{\mbox{{\it \boldmath$#1$}}}

\lefthead{Hoekstra et al.}
\righthead{Masses for moderate luminosity X-ray clusters}

\begin{document}

\title{The mass-$L_x$ relation for moderate luminosity X-ray clusters\footnotemark[$\dagger$]}

\footnotetext[$\dagger$]{Based on observations obtained with HST}

\author{Henk~Hoekstra\altaffilmark{1,2}, Megan~Donahue\altaffilmark{3}, 
Christopher~J.~Conselice\altaffilmark{4}, 
Brian~R.~McNamara\altaffilmark{5,6,7}, 
G. Mark Voit\altaffilmark{3}}

\altaffiltext{1}{Leiden Observatory, Leiden University, Leiden, Netherlands} 
\altaffiltext{2}{Department of Physics and Astronomy, University of
Victoria, Victoria, BC, Canada}
\altaffiltext{3}{Physics and Astronomy Department, Michigan State University, East Lansing, MI} 
\altaffiltext{4}{University of Nottingham, England} 
\altaffiltext{5}{Department of Physics and Astronomy, University of Waterloo, Waterloo, ON, Canada}
\altaffiltext{6}{Perimeter Institute for Theoretical Physics, Waterloo, ON, Canada}
\altaffiltext{7}{Harvard-Smithsonian Center for Astrophysics, Cambridge, MA}

\begin{abstract}

We present measurements of the masses of a sample of 25 moderate X-ray
luminosity clusters of galaxies from the 160 square degree ROSAT
survey. The masses were obtained from a weak lensing analysis of deep
$F814W$ images obtained using the Advanced Camera for Surveys
(ACS). We present an accurate empirical correction for the effect of
charge transfer (in)efficiency on the shapes of faint galaxies. A
significant lensing signal is detected around most of the
clusters. The lensing mass correlates tightly with the cluster
richness. We measured the intrinsic scatter in the scaling relation
between $M_{2500}$ and $L_X$ to be $\sigma_{\log
  {L_X|M}}=0.23^{+0.10}_{-0.04}$. The best fit power law slope and
normalisation are found to be $\alpha=0.68\pm0.07$ and
$M_X=(1.2\pm0.12)\times h_{70}^{-1} 10^{14}\msun$ (for $L_X=2\times
10^{44} h_{70}^{-2}$ erg/s). These results agree well with a number of
recent studies, but the normalisation is lower compared to the study
of \cite{Rykoff08b}. One explanation for this difference may be the
fact that (sub)structures projected along the line-of-sight boost both
the galaxy counts and the lensing mass. Such superpositions lead to an
increased mass at a given $L_X$ when clusters are binned by richness.

\end{abstract}


\section{Introduction}

Clusters of galaxies are known to exhibit correlations between their
various observable properties, such as the well-known relation between
X-ray luminosity and X-ray temperature. These scaling relations are
the result of the various processes that govern the formation and
subsequent evolution of galaxy clusters. Hence, the study of cluster
scaling laws provides the basis of testing models for the formation of
clusters of galaxies and of galaxies themselves.  For instance, N-body
codes can nowadays reliably and robustly predict the evolution of the
mass function of cluster halos \citep[e.g.,][]{Evrard02}. Hydrodynamic
simulations and semi-analytic techniques then allow us to predict the
X-ray and optical appearance of these halos \citep[e.g.,][]{Voit05,Nagai07,
Bower08}.

The mean relation and the intrinsic dispersion of the resulting
mass-observable relations provide a test of the adequacy of the
physical processes included in the simulations: a realistic simulation
should reproduce the mean relation, the intrinsic scatter, and its
evolution (if any). Such comparisons have recently led to the
realization that cluster simulations must include radiative cooling
and feedback from supernovae and AGNs in order to successfully explain
the observed scaling relations (see \cite{Voit05} for a review).  While
the core regions of clusters remain difficult to model accurately,
high-resolution simulations with cooling and feedback now produce
clusters whose X-ray properties agree well with those of observed
clusters outside the central 100 kpc or so \citep{Nagai07}.

An empirically determined mass scaling relation is not only useful to
directly test our understanding of cluster physics, it also allows one
to relate the observables directly to the mass of cluster-sized halos.
The latter, for instance, enables one to transform the observed
luminosity function into a mass function, which in turn can be
compared to different cosmological models. Such studies have already
provided interesting constraints on cosmological parameters which
complement constraints from other probes \citep[e.g.,][]{Henry00,
Borgani01,Reiprich02,Henry04,Gladders07,Henry09,Vikhlinin09b,Mantz10a}.

To determine cluster masses, a number of methods are available.
Dynamical techniques, such as the assumption of hydrostatic
equilibrium in X-ray observations, have been widely used. The inferred
masses, however, are likely to be systematically underestimated by up
to $20\%$ because of pressure support from turbulence and energy input
from active galactic nuclei \citep[e.g.,][]{Evrard96,Rasia06,Nagai07}.
Masses derived from weak gravitational lensing do not require
assumptions about the dynamical state of the cluster and can in
principle be compared directly to the outcome of numerical
simulations. It is worth noting, however, that lensing is sensitive to
all structure along the line-of-sight. Weak lensing is now a
well-established technique \citep[for a review see][]{HJ08} and recent
work has indeed provided support for non-hydrostatic gas in the
outskirts of X-ray luminous clusters of galaxies \citep{Mahdavi08}.

To date, most work has focussed on the most massive clusters at
intermediate redshifts \citep[e.g.,][]{Bardeau07,Hoekstra07,Zhang08},
because of the relatively large lensing signal. In this paper we focus
on extending the mass range towards clusters with lower X-ray
luminosities ($L_x < {\rm few~}10^{44}$~erg/s). To detect the lensing
signal of these lower mass clusters, the number density of lensed
background galaxies needs to be increased significantly, compared to
deep ground based observations.  The reason for this requirement is
that the error in the weak lensing mass estimate is set by the
intrinsic shapes of the galaxies and their number density.

To this end, we have conducted a weak-lensing analysis on images
obtained during a snapshot survey of 25 moderate luminosity
X-ray clusters with the Advanced Camera for Surveys (ACS) on board the
Hubble Space Telescope (HST).  These clusters were randomly selected
from the ROSAT 160 Square Degree (160SD) Cluster Catalog
\citep{Mullis03, Vikhlinin98}. The clusters in the 160SD sample were
serendipitously discovered in pointed, relatively wide-field (30'
diameter) observations of the Position-Sensitive Proportional Counter
(PSPC) on board the ROSAT X-ray observatory (1990-1999). Seventy-two
X-ray clusters between $z=0.3-0.7$ were discovered (see
Fig.~\ref{sample}). All have been identified and assigned redshifts
\citep{Mullis03}. Because this survey included pointed observations
that were quite long, some of these clusters are among the faintest
clusters of galaxies known at these moderately high
redshifts. Therefore, our new weak lensing measurements extend the
X-ray luminosity limit of the mass-$L_x$ relation by almost an order
of magnitude, based on targeted observations. We note that studies of
the ensemble-averaged properties of clusters discovered in the SDSS
\citep{Rykoff08b} and X-ray groups in COSMOS \citep{Leauthaud10} have
pushed the limits to even lower luminosities.

The structure of the paper follows. In \S2 we describe our data and
weak lensing analysis. In particular we discuss how we correct for the
effects of CTE in our ACS data and how we correct for PSF
anisotropy. The measurements of the cluster masses, and the comparison
to the X-ray properties are presented in \S3. We compare our results
to previous work and examine biases in our mass estimates that arise
from uncertainties in the position of the cluster center and sample
selection. Throughout this paper we assume a flat $\Lambda$CDM
cosmology with $\Omega_m=0.3$ and $H_0=70h_{70}$~km/s/Mpc.

\begin{table*}
\begin{center}
\caption{Summary of the observational data for the cluster sample
\label{tabsample}}
\begin{tabular}{crccccccccc}
\hline
\hline
name &  number & $z$ & $ L_X$ & RA$_{\rm X-ray}$ & DEC$_{\rm X-ray}$ & RA$_{\rm BCG}$  & DEC$_{\rm BCG}$ & $Q_{\rm BCG}$ & $\langle\beta\rangle$ & $\langle\beta^2\rangle$\\
 (1) &    (2)    &  (3)           & (4) & (5)             &  (6)            & (7)   & (8)                   & (9) & (10) & (11) \\
\hline
RXJ$0056.9-2740$ & 6    & 0.563 & 1.14 & ~$00^h56^m56.1^s$ & $-27^\circ40'12''$ & ~$00^h56^m56.98^s$ & $-27^\circ40'29.9''$ & 2 & 0.42 & 0.23 \\
RXJ$0110.3+1938$ & 8    & 0.317 & 0.40 & ~$01^h10^m18.0^s$ & $+19^\circ38'23''$ & ~$01^h10^m18.22^s$ & $+19^\circ38'19.4''$ & 0 & 0.61 & 0.43 \\
RXJ$0154.2-5937$ & 20   & 0.360 & 0.90 & ~$01^h54^m14.8^s$ & $-59^\circ37'48''$ & ~$01^h54^m13.72^s$ & $-59^\circ37'31.0''$ & 1 & 0.57 & 0.38 \\
RXJ$0522.2-3625$ & 41   & 0.472 & 2.07 & ~$05^h22^m14.2^s$ & $-36^\circ25'04''$ & ~$05^h22^m15.48^s$ & $-36^\circ24'56.1''$ & 1 & 0.48 & 0.29 \\
RXJ$0826.1+2625$ & 52   & 0.351 & 0.65 & ~$08^h26^m06.4^s$ & $+26^\circ25'47''$ & ~$08^h26^m09.45^s$ & $+26^\circ25'03.1''$ & 0 & 0.58 & 0.39 \\
RXJ$0841.1+6422$ & 56   & 0.342 & 1.60 & ~$08^h41^m07.4^s$ & $+64^\circ22'43''$ & ~$08^h41^m07.65^s$ & $+64^\circ22'26.0''$ & 2 & 0.59 & 0.40 \\
RXJ$0847.1+3449$ & 59   & 0.560 & 1.17 & ~$08^h47^m11.3^s$ & $+34^\circ49'16''$ & ~$08^h47^m11.79^s$ & $+34^\circ48'51.8''$ & 1 & 0.42 & 0.23 \\
RXJ$0910.6+4248$ & 69  & 0.576 & 1.44 & ~$09^h10^m39.7^s$ & $+42^\circ48'41''$ & ~$09^h10^m40.53^s$ & $+42^\circ49'59.1''$ & -1&0.41 & 0.23 \\
RXJ$0921.2+4528$ & 70  & 0.315 & 1.40 & ~$09^h21^m13.4^s$ & $+45^\circ28'50''$ & ~$09^h21^m13.46^s$ & $+45^\circ28'56.1''$ & 1 & 0.61 & 0.43 \\
RXJ$0926.6+1242$ & 71  & 0.489 & 2.04 & ~$09^h26^m36.6^s$ & $+12^\circ42'56''$ & ~$09^h26^m36.70^s$ & $+12^\circ43'03.8''$ & 2 & 0.47 & 0.28 \\
RXJ$0957.8+6534$ & 80  & 0.530 & 1.39 & ~$09^h57^m53.2^s$ & $+65^\circ34'30''$ & ~$09^h57^m51.22^s$ & $+65^\circ34'25.1''$ & 2 & 0.44 & 0.25 \\
RXJ$1015.1+4931$ & 88  & 0.383 & 0.78 & ~$10^h15^m08.5^s$ & $+49^\circ31'32''$ & ~$10^h15^m08.44^s$ & $+49^\circ31'50.8''$ & 2 & 0.55 & 0.36 \\
RXJ$1117.2+1744$ & 96  & 0.305 & 0.54 & ~$11^h17^m12.0^s$ & $+17^\circ44'24''$ & ~$11^h17^m11.23^s$ & $+17^\circ44'00.5''$ & 2 & 0.62 & 0.44 \\
RXJ$1117.4+0743$ & 97  & 0.477 & 0.76 & ~$11^h17^m26.1^s$ & $+07^\circ43'35''$ & ~$11^h17^m26.04^s$ & $+07^\circ43'38.3''$ & 1 & 0.48 & 0.29 \\
RXJ$1123.1+1409$ & 101 & 0.340 & 1.01 & ~$11^h23^m10.2^s$ & $+14^\circ09'44''$ & ~$11^h23^m10.95^s$ & $+14^\circ08'36.4''$ & 1 & 0.59 & 0.40 \\
RXJ$1354.2-0221$ & 151 & 0.546 & 1.53 & ~$13^h54^m16.9^s$ & $-02^\circ21'47''$ & ~$13^h54^m17.19^s$ & $-02^\circ21'59.2''$ & 2 & 0.43 & 0.24 \\
RXJ$1524.6+0957$ & 170 & 0.516 & 3.94 & ~$15^h24^m40.3^s$ & $+09^\circ57'39''$ & ~$15^h24^m41.56^s$ & $+09^\circ57'34.3''$ & 1 & 0.45 & 0.26 \\
RXJ$1540.8+1445$ & 172 & 0.441 & 0.81 & ~$15^h40^m53.3^s$ & $+14^\circ45'34''$ & ~$15^h40^m53.96^s$ & $+14^\circ45'56.0''$ & 1 & 0.51 & 0.31 \\
RXJ$1642.6+3935$ & 186 & 0.355 & 0.62 & ~$16^h42^m38.9^s$ & $+39^\circ35'53''$ & ~$16^h42^m38.35^s$ & $+39^\circ36'10.4''$ & 1 & 0.58 & 0.39 \\
RXJ$2059.9-4245$ & 199 & 0.323 & 2.44 & ~$20^h59^m55.2^s$ & $-42^\circ45'33''$ & ~$20^h59^m54.92^s$ & $-42^\circ45'32.1''$ & 2 & 0.61 & 0.42 \\
RXJ$2108.8-0516$ & 200 & 0.319 & 0.81 & ~$21^h08^m51.2^s$ & $-05^\circ16'49''$ & ~$21^h08^m51.17^s$ & $-05^\circ16'58.4''$ & 2 & 0.61 & 0.42 \\
RXJ$2139.9-4305$ & 203 & 0.376 & 0.59 & ~$21^h39^m58.5^s$ & $-43^\circ05'14''$ & ~$21^h39^m58.22^s$ & $-43^\circ05'13.9''$ & 2 & 0.56 & 0.37 \\
RXJ$2146.0+0423$ & 204 & 0.531 & 2.61 & ~$21^h46^m04.8^s$ & $+04^\circ23'19''$ & ~$21^h46^m05.52^s$ & $+04^\circ23'14.3''$ & 2 & 0.44 & 0.25 \\
RXJ$2202.7-1902$ & 205 & 0.438 & 0.70 & ~$22^h02^m44.9^s$ & $-19^\circ02'10''$ & ~$22^h02^m45.50^s$ & $-19^\circ02'21.1''$ & 2 & 0.51 & 0.31 \\
RXJ$2328.8+1453$ & 219 & 0.497 & 1.02 & ~$23^h28^m49.9^s$ & $+14^\circ53'12''$ & ~$23^h28^m52.27^s$ & $+14^\circ52'42.8''$ & 1 & 0.46 & 0.27 \\
\hline
\hline
\end{tabular}
\tablecomments{(2) entry number for the cluster in \cite{Mullis03};
  (3) cluster redshift from \cite{Mullis03}; (4) restframe X-ray
  luminosity in the $0.1-2.4$ keV band in units of
  $10^{44}h_{70}^{-2}$ ergs/s.  (5) \& (6) location (J2000.0) of the
  X-ray position from \cite{Mullis03}. (7) \& (8) location (J2000.0)
  of the assumed brightest cluster galaxy; (9) `quality' of BCG
  identification (see \S2.5 for details); (10) \& (11) mean
  $\beta=D_{ls}/D_s$ and variance as explained in the text.}
\end{center}
\end{table*}

~

\section{Data and analysis}

The data studied here were obtained as part of a snapshot program (PI:
Donahue) to study a sample of clusters found in the 160 square degree
survey \citep{Vikhlinin98,Mullis03}. The clusters from the latter
survey were selected based on the serendipitous detection of extended
X-ray emission in ROSAT PSPC observations, resulting in a total survey
area of 160 deg$^2$. A detailed discussion of the survey can be found
in \cite{Vikhlinin98}. The sample was reanalysed by \cite{Mullis03},
which also lists spectroscopic redshifts for most of the clusters.

\begin{figure}
\begin{center}
\leavevmode
\hbox{%
\epsfxsize=8.5cm
\epsffile{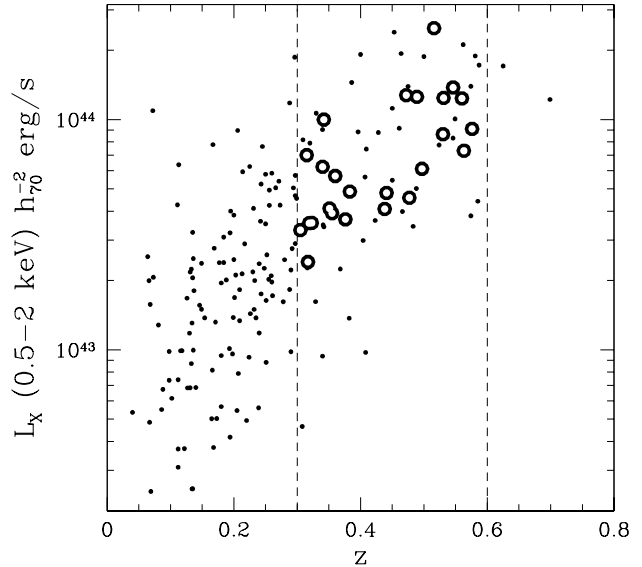}}
\caption{X-ray luminosity as a function of redshift for the 160 square
degree survey from \cite{Mullis03}. The large open points
indicate the clusters for which HST imaging data were obtained.
\label{sample}}
\end{center}
\end{figure}

The X-ray luminosity as a function of cluster redshift is plotted in
Figure~\ref{sample}. The HST snapshot program targeted clusters with
$0.3<z<0.6$, which were observed during HST cycles 13 and 14. This
resulted in a final sample of 25 clusters that were imaged in the
$F814W$ filter with integration times of $\sim 2200$s. Note that these
were drawn randomly from the sample, and therefore represent a fair
sample of an X-ray flux limited survey. The clusters with HST data are
indicated in Figure~\ref{sample} by the large open points.

Table~\ref{tabsample} provides a list of the clusters that were
observed. It also lists the restframe X-ray luminosities in the
$0.1-2.4$ keV band. The values were taken from the BAX X-Rays Galaxy
Clusters Database\footnote{http://bax.ast.obs-mip.fr/} and converted
to the cosmology we adopted here.  As explained on the BAX website,
the luminosities were derived from the flux measurements using a
Raymond-Smith type spectrum assuming a metallicity of 0.33 times
solar.

Each set of observations consists of four exposures, which allow for
efficient removal of cosmic rays. We use the {\tt multidrizzle} task
to remove the geometric distortion of ACS images and to combine the
multiple exposures \citep{Koekemoer02}.  This task also removes cosmic
rays and bad pixels, etc. Because of the large geometric distortion
and offsets between the individual exposures, the images need to be
resampled before co-addition. A number of options have been studied by
\cite{Rhodes07} who suggest to use a Gaussian kernel and an output
pixel scale of 0.03".  However, this procedure leads to correlated
noise in the final image. Instead we opt for the {\tt lanczos3}
resampling kernel and keep the original pixel size of 0.05". These
choices preserve the noise properties much better, while the images of
the galaxies are adequately sampled. These co-added images are
subsequently used in the weak lensing analysis.

\subsection{Shape analysis}

The first step in the weak lensing analysis is the detection of stars
and galaxies, for which we use SExtractor \citep{Bertin96}. The next
important step is the unbiased measurement of the shapes of the faint
background galaxies that will be used to quantify the lensing
signal. To do so, we measure weighted quadrupole moments, defined as

\begin{equation}
I_{ij}=\int d^2{\bma x} x_i x_j W({\bma x}) f({\bma x}),
\end{equation}

\noindent where $W({\bma x})$ is a Gaussian with a dispersion $r_g$,
which is matched to the size of the object \citep[see][for more
  details]{KSB95, Hoekstra98}. These quadrupole moments are combined to
form the two-component polarization

\begin{equation}
e_1=\frac{I_{11}-I_{22}} {I_{11}+I_{22}},
~{\rm and}~e_2=\frac{I_{12}}{I_{11}+I_{22}}.
\end{equation}

However, we cannot simply use the observed polarizations, because they
have been modified by a number of instrumental effects. Although the
PSF of HST is small compared to ground based data, the shapes of the
galaxies are nonetheless slightly circularized, lowering the amplitude
of the observed lensing signal. Furthermore, the PSF is not round and
the resulting anisotropy in the galaxy shapes needs to be undone. To
do so, we use the well established method proposed by \cite{KSB95},
with the modifications provided in \cite{LK97} and
\cite{Hoekstra98,Hoekstra00}.

\subsection{Correction for CTE degradation}

An issue that is relevant for ACS observations is the degradation of
the charge transfer efficiency (CTE) with time. Over time, cosmic rays
cause an increasing number of defects in the detector. During
read-out, these defects can trap charges for a while. The delayed
transfer of charge leads to a trail of electrons in the read-out
direction, causing objects to appear elongated. The effect is
strongest for faint objects, because brighter objects quickly fill the
traps.  \cite{Rhodes07} provide a detailed discussion of CTE effects
and their impact on weak lensing studies. Similar to \cite{Rhodes07}
we derive an empirical correction for CTE, but our adopted approach
differs in a number of ways.

The presence of CTE in our data leads to a slight modification of our
lensing analysis. First, we note that CTE affects only $e_1$ and that
the change in shape occurs during the read-out stage. Hence the
measured value for $e_1$ needs to be corrected for CTE {\it before}
the correction for PSF anisotropy and circularization:

\begin{equation}
e_1=e_1^{\rm obs} - e_1^{\rm CTE},
\end{equation}

\noindent where $e_1^{\rm CTE}$ is the predicted change in
polarization given by the model derived in the Appendix. To derive our
CTE model, we used observations of the star cluster NGC~104 as well as
100 exposures from the COSMOS survey \citep{Scoville07}. The
observations of a star cluster allow us to study the effect of CTE as
a function of position, time and signal-to-noise ratio with high
precision, because stars are intrinsically round after correction for
PSF anisotropy. We find that the CTE effect increases linearly with
time and distance from the read-out electronics. The amplitude of the
effect is observed to be proportional to $\sqrt{S/N}$. We use 100
exposures from COSMOS to examine how the CTE effect depends on source
size. We find a strong size dependence, with $e_1^{\rm CTE}\propto
r_g^{-2}$ (where $r_g$ is the dispersion of the Gaussian weight
function used to measure the quadrupole moments).

\subsection{Correction for the PSF} 

Once the CTE effect has been subtracted for all objects (including
stars), the lensing analysis proceeds as described in
\cite{Hoekstra98}. Hence the next step is to correct both polarization
components for PSF anisotropy. The ACS PSF is time dependent and
therefore a different PSF model is required for each observation.  An
added complication is the fact that only a limited number of stars are
observed in each image.

To estimate the spatial variation of the PSF anisotropy a number of
procedures have been proposed
\citep{Rhodes07,Schrabback07,Schrabback10}. Here we opt for a simple
approach, similar to the one used in \cite{Hoekstra98}.

\begin{figure}
\begin{center}
\leavevmode
\hbox{%
\epsfxsize=8.5cm
\epsffile{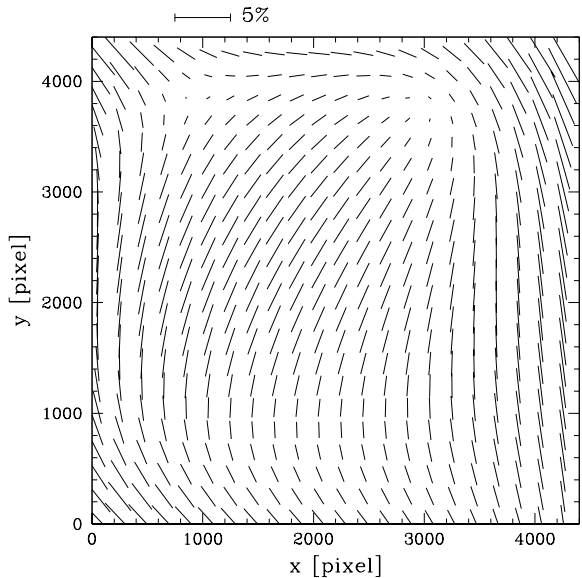}}
\caption{PSF reference model based on observations of NGC104. The
sticks show the direction and amplitude (indicated by the length) of
the polarization of the PSF. The PSF anisotropy is generally small,
but can reach 4\% at the edges of the field.
\label{psfmod}}
\end{center}
\end{figure}

We use observations of the star cluster NGC104 to derive an adequate
model for PSF anisotropy. These data were taken at the start of ACS
operations (PID 9018), and therefore do not suffer from CTE effects.
We model the PSF anisotropy by a third order polynomial in both $x$
and $y$. Such a model would not be well constrained by our galaxy cluster
data, but it is thanks to the high number density of stars in NGC104.
We select one of the models as our reference, because the PSF pattern
varied only mildly from exposure to exposure. The resulting reference
model is shown in Figure~\ref{psfmod}. The PSF anisotropy is fairly
small, but can reach $\sim 4\%$ towards the edges of the field.

Most of the variation in the ACS PSF arises from focus changes, which
are the result of changes in the telescope temperature as it orbits
the Earth. We therefore expect that a scaled version of our model can
capture much of the spatial variation: as the detector moves from one
side of the focus to the other, the direction of PSF anisotropy
changes by 90 degrees, which corresponds to a change of sign in the
polarization.  To account for additional low order changes, we also
include a first order polynomial:

\begin{equation}
e_i^{\rm PSF}=\alpha e_i^{\rm NGC104} + a_0 +a_1 x +a_2 y.
\end{equation}

\begin{figure}
\begin{center}
\leavevmode
\hbox{%
\epsfxsize=8.5cm
\epsffile{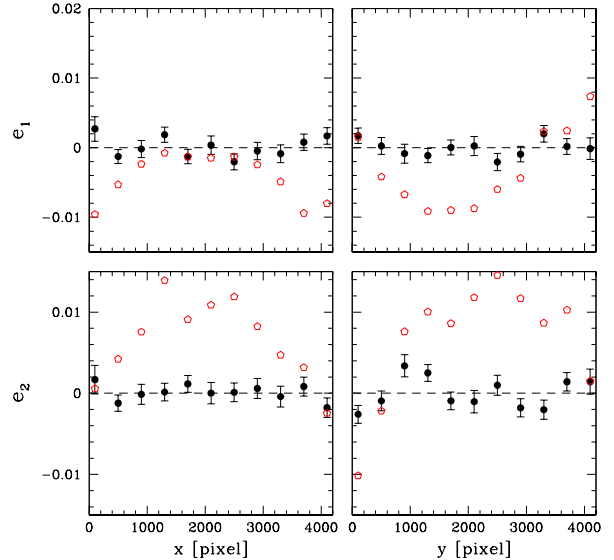}}
\caption{The ensemble averaged PSF polarization $e_1$ and
$e_2$ as a function of position using stars detected in our galaxy
cluster exposures. The open circles show the PSF anisotropy
before correction, whereas the solid points correspond to
the residuals after applying the best fit PSF model to the
stars.
\label{psfres}}
\end{center}
\end{figure}

This simple model, with only 4 parameters, is used to characterize the
PSF anisotropy for each individual galaxy cluster exposure. The number of
stars in the galaxy cluster images ranges from 7 to 84, with a median of
20. To examine how well our approach works, we averaged the shapes of
all the stars in our 25 galaxy cluster exposures. The ensemble averaged PSF
polarization as a function of position is shown in
Figure~\ref{psfres}. The open points show the average PSF shapes
before correction. More importantly, the filled points show the
residuals in polarization after the best fit model for each galaxy cluster
pointing has been subtracted. These results suggest our model is
adequate for our weak lensing analysis.

\subsection{Lensing signal}

The corrected galaxy shapes are used to measure the weak lensing
signal. A convenient way to quantify the lensing signal is by
computing the azimuthally averaged tangential shear $\gamma_T$ as a
function of radius from the cluster center (see \S2.5 for our choice
of the center). It can be related to the surface density through

\begin{equation}
\langle\gamma_t\rangle(r)=\frac{\bar\Sigma(<r) - \bar\Sigma(r)}
{\Sigma_{\rm crit}}=\bar\kappa(<r)-\bar\kappa(r),
\end{equation}

\noindent where $\bar\Sigma(<r)$ is the mean surface density within an
aperture of radius $r$, and $\bar\Sigma(r)$ is the mean surface
density on a circle of radius $r$. The convergence $\kappa$, or
dimensionless surface density, is the ratio of the surface density and
the critical surface density $\Sigma_{\rm crit}$, which is given by

\begin{equation}
\Sigma_{\rm crit}=\frac{c^2}{4\pi G}\frac{D_s}{D_l D_{ls}},
\end{equation}

\noindent where $D_l$ is the angular diameter distance to the
lens. $D_{s}$ and $D_{ls}$ are the angular diameter distances from the
observer to the source and from the lens to the source,
respectively. The parameter $\beta=\max[0,D_{ls}/D_s]$ is a measure of
how the amplitude of the lensing signal depends on the redshifts of
the source galaxies (where a value of 0 is assigned to sources with a
redshift lower than that of the cluster).

One of the advantages of weak lensing is that the (projected) mass can
be determined in a model-independent way. To derive accurate masses
requires wide field imaging data, which we lack because of the small
field of view of ACS. Instead we fit parametric models to the lensing
signal.

The singular isothermal sphere (SIS) is a simple model to describe the
cluster mass distribution. In this case the convergence and tangential
shear are equal:

\begin{equation}
\kappa(r)=\gamma_T(r)=\frac{r_E}{2r},
\end{equation}

\noindent where $r_E$ is the Einstein radius. In practice we do not
observe the tangential shear directly, but we observe the reduced
shear $g_T$ instead:

\begin{equation}
g_T=\frac{\gamma_T}{1-\kappa}.
\end{equation}

\noindent We correct our model predictions for this effect. Under the
assumption of isotropic orbits and spherical symmetry, the Einstein
radius (in radian) can be expressed in terms of the line-of-sight
velocity dispersion:

\begin{equation}
r_E=4\pi \left(\frac{\sigma}{c}\right)^2 \beta.
\end{equation}

\noindent We use this model when listing lensing inferred velocity
dispersions in Table~\ref{tabresult}.

A more physically motivated model is the one proposed by \cite{NFW97}
who provide a fitting function to the density profiles observed in
numerical simulations of cold dark matter. We follow the conventions
outlined in \cite{Hoekstra07}, but use an updated relation between the
virial mass $M_{\rm vir}$ and concentration $c$. \cite{Duffy08}
studied numerical simulations using the best fit parameters of the
WMAP5 cosmology \citep{Komatsu09}. The best fit $c(M_{\rm vir})$ is given by:

\begin{equation}
c=7.85\left({\frac{M_{\rm vir}}{2\times 10^{12}}}\right)^{-0.081}{(1+z)^{-0.71}}.\label{mcrel}
\end{equation}

\noindent We use this relation when fitting the NFW model to the
lensing signal. Rather than the virial mass, in Table~\ref{tabresult}
we list $M_{2500}$ which is the mass within a radius $r_{2500}$ where
the mean mass density of the halo is 2500 times the critical density
at the redshift of the cluster. We note that a different choice for
the mass-concentration relation will change the inferred masses much.
We found that the inferred values for $M_{2500}$ change by $5-10\%$ if
we change the pre-factor in Equation~\ref{mcrel} by $\pm 1$, which is
much smaller than our statistical uncertainties. When comparing to
ensemble averaged results from other studies, however, differences in
the adopted mass-concentration relation may become a dominant source
of uncertainty.

Finally, we note that the lensing signal is sensitive to all matter
along the line-of-sight. As shown in \cite{Hoekstra01,Hoekstra03}, the
large-structure in the universe introduces cosmic noise, which
increases the formal error in the mass estimate, compared to just the
statistical uncertainty. The listed uncertainties in the weak lensing
masses include this contribution.

\subsection{Cluster center}

We have to choose a position around which we measure the weak lensing
signal as a function of radius. An offset between the adopted position
and the `true' center of the dark matter halo will lead to an
underestimate of the cluster mass. Possible substructure in the
cluster core complicates a simple definition of the cluster center,
but it can also lead to biased mass estimates
\citep[e.g.,][]{Hoekstra02}.  We expect our results to be less
affected by substructure because the ACS data used here extend to
larger radii than the WFPC2 observations discussed in \cite{Hoekstra02}.

The resulting bias depends on the detailed procedure that is used to
to interpret the lensing signal. For instance, \cite{Hoekstra07} used
wide field imaging data to measure the lensing signal out to large
radii and to derive (almost) model-independent masses. This procedure
minimizes the effect of centroiding errors because the large-scale
signal is affected much less, compared to the signal on small
scales. Unfortunately, the ACS field of view is relatively small,
which prevents us from following the same approach and we fit a
parameterized mass model to the lensing signal instead.  To reduce the
sensitivity of our results to centroiding errors and central
substructure, we fit the NFW and SIS models to the tangential
distortion at radii $200<r<750 h_{70}^{-1}$kpc.

\begin{figure}
\begin{center}
\leavevmode
\hbox{%
\epsfxsize=8.5cm
\epsffile{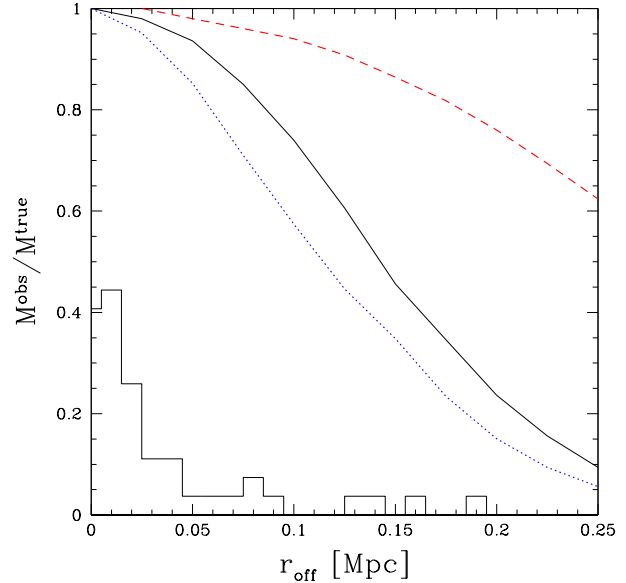}}
\caption{Plot of the ratio of the inferred lensing mass and the true
mass as a function of centroid offset. The lensing mass is obtained by
fitting an NFW model to the shear at $200-750 h^{-1}_{70}$ kpc (solid
black curve) and $0.5-1.5h^{-1}_{70}$Mpc (dashed red curve). The blue
dotted line corresponds to the bias if all data within $750
h^{-1}_{70}$ kpc are used. The results presented here are for a
cluster with a mass $M_{2500}=2\times 10^{14}h_{70}^{-1} \msun$, but
we note that the bias varies only by a few percent over the range of
masses we consider here. The histogram indicates the frequency of 
offsets found for massive clusters from \cite{Bildfell08}.
\label{bias_offset}}
\end{center}
\end{figure}

The advantage of restricting the analysis to larger scales is also
evident from Figure~\ref{bias_offset}, where we show the bias in mass
as a function of centroid offset $r_{\rm off}$. The solid curve shows
the results when fitting an NFW model to the signal within $200-750
h^{-1}_{70}$ kpc, which is the range we use for the ACS data studied
here.

\cite{Mullis03} provide cluster positions based on the X-ray
emission. For the clusters in our sample, the listed centroiding
uncertainty depends on the X-ray luminosity (with smaller values for
the more luminous systems). An alternative approach is to use the
location of the brightest cluster galaxy (BCG). The resulting
positions are listed in Table~\ref{tabsample}. In many cases a clear
candidate can be identified (indicated by a value of $Q_{\rm BCG}$ of
1 or 2 in Table~\ref{tabsample}), but in a number of cases the choice
is ambiguous (indicated by $Q_{\rm BCG}$ -1 or 0).

High quality X-ray data can be used to refine the centers, but such
data are lacking for our sample. Even if such data were available,
there can be a offset between the BCG and the peak in the X-ray
emission \citep[e.g.,][]{Bildfell08}. The distribution of offsets
observed by \cite{Bildfell08} for the massive clusters in the
Canadian Cluster Comparison Project, are indicated by the histogram in
Figure~\ref{bias_offset}. It is clear that the offsets are typically
less than 50 kpc, leading to biases less than 5\%. The larger offsets
are found for merging massive systems where the identification of the
BCG is difficult. Such major mergers do not appear to be present in
the sample of clusters studied here.

We compared the offset between the X-ray position and the adopted
location of the BCG to the uncertainty listed in \cite{Mullis03}
and find fair agreement: most of the BCGs are located within the
radius corresponding to the 90\% confidence X-ray position error
circle. Only for 4 low luminosity systems do we find an offset that is
much larger. The positions of the BCGs can be determined more
accurately, and we therefore adopt these as the cluster centers when
listing our mass estimates.

\subsection{Source galaxies}

The lensing signal is largest for background galaxies at redshifts
much larger than the cluster. We lack redshift information for our
sources and instead we select a sample of faint (distant) galaxies
with $24<F814W<26.5$, which also reduces contamination by cluster
members which dominate the counts at bright magnitudes. Nonetheless,
contamination by cluster members cannot be ignored because we lack
color information. We note that adding a single color would not
improve the situation significantly because the faint members span a
wide range in color, unlike the bright members, almost all of which
occupy a narrow red-sequence.

\begin{figure}
\begin{center}
\leavevmode
\hbox{%
\epsfxsize=8.5cm
\epsffile{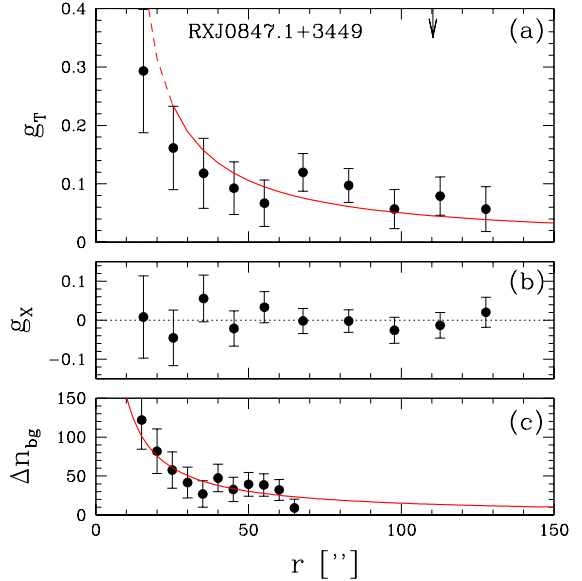}}
\caption{{\it panel a}: Tangential distortion as a function of radius
for RXJ0847.1+3449. The measurements have been corrected for
contamination by faint cluster member using the best fit model to the
excess counts shown in panel~c. The best fit SIS model based on a fit
to the measurements at $r>25$ arcseconds is indicated by the drawn
line.  The arrow indicates a radius of $0.5h^{-1}$Mpc. {\it panel b}:
The signal when the sources are rotated by 45 degrees. In this case
the signal should vanish if the measurements shown in panel~a are due
to gravitational lensing. {\it panel c}: Excess counts of galaxies
with $24<F814W<26.5$ as compared to the mean counts based on 100
frames from the COSMOS survey. The expected background counts are
$\sim 61$ arcmin$^{-2}$. The drawn line is the best fit $1/r$ model,
which is used to correct the measurements of the tangential distortion.
\label{gtprof}}
\end{center}
\end{figure}

To account for contamination by cluster members we measure the number
density of faint galaxies as a function of distance from the adopted
cluster center and boost the signal by the inferred fraction of
cluster members. The corrected tangential distortion for
RXJ0847.1+3449 as a function of distance from the adopted cluster
centre is shown in Figure~\ref{gtprof}a. The red line shows the best
fit singular isothermal sphere model (fitted to radii $>25''$). If the
observed signal is caused by gravitational lensing, the signal should
vanish if the source galaxies are rotated by 45 degrees.
Figure~\ref{gtprof}b shows the results of this test, which are indeed
consistent with no signal.

The lower panel in Figure~\ref{gtprof} shows the excess counts $\Delta
n_{\rm bg}$ as a function of radius for the cluster RXJ0847.1+3449
($z=0.56$).  To determine the background count levels we used the 100
COSMOS pointings that were used to measure the size-dependence of
CTE. We measure the excess counts for each cluster individually,
because the sample spans a fair range in redshift and mass.

This cluster shows a signicant excess of faint galaxies
over the background number density of $\sim 61$ arcmin$^{-2}$.  We
assume that these faint cluster members are oriented randomly and thus
simply dillute the lensing signal. To quantify the overdensity of
faint members we adopt $\Delta n_{\rm bg}\propto r^{-1}$ and determine
the best fit normalization for each cluster separately (indicated by
the drawn line in Figure~\ref{gtprof}c. This simple model is
used to correct the observed tangential distortion for contamination
by cluster members. We find that this correction leads to an average
increase in the best fit Einstein radius of $\sim 20\%$.  The
uncertainty in the amplitude of the contamination is included in our
quoted measurement errors (which are increased by less than 5\%).

The conversion of the lensing signal into an estimate for the mass of
the cluster requires knowledge of the redshift distribution of the
source galaxies. These are too faint to be included in spectroscopic
redshift surveys or even from ground based photometric redshift
surveys \citep[e.g.,][]{Wolf04,Ilbert06}. Instead we use the
photometric redshift distributions determined from the Hubble Deep
Fields \citep{HDF}. For the range in apparant magnitudes used in the
lensing analysis, the resulting average value for
$\langle\beta\rangle$ and the variance $\langle\beta^2\rangle$ are
listed in Table~\ref{tabsample}. As discussed in \cite{Hoekstra00} the
latter quantity is needed to account for the fact that we measure the
reduced shear. We estimate the uncertainty in $\beta$ by considering
the variation between the two HDFs. As the average source redshift is
much higher than the cluster redshifts, the resulting relative
uncertainties are small: $\sim 2\%$ at $z=0.3$, increasing to $\sim
5\%$ at $z=0.6$, much smaller than our statistical errors.

\begin{table*}
\begin{center}
\caption{Measurements of cluster properties
\label{tabresult}}
\begin{tabular}{llllllll}
\hline
\hline
name\hspace{2.5cm} & $r_E$\hspace{1.5cm} & $\sigma$\hspace{1.5cm} & $r_{2500}$\hspace{1cm} & $M_{2500}$ & $N_{2500}$\\
     &               ['']  & [km/s]   & [$h_{70}^{-1}$Mpc] & [$h_{70}^{-1}10^{13}$M$_\odot$] & \\
\hline
RXJ$0056.9-2740$ & $5.4\pm1.5$ & $678^{+88}_{-101}$ &	0.270 & $5.2^{+4.2}_{-3.0}$ & $13\pm4$ \\
RXJ$0110.3+1938$ & $5.8\pm1.3$ & $577^{+62}_{-69}$ &	0.293 & $5.0^{+3.4}_{-2.6}$ & $4\pm2$ \\
RXJ$0154.2-5937$ & $3.6\pm1.4$ & $474^{+83}_{-101}$ &	0.218 & $2.2^{+2.6}_{-1.4}$ & $1\pm1$ \\
RXJ$0522.2-3625$ & $6.8\pm1.3$ & $710^{+66}_{-73}$ &	0.313 & $7.2^{+4.2}_{-3.1}$ & $7\pm3$ \\
RXJ$0826.1+2625$ & $2.4\pm1.8$ & $384^{+124}_{-192}$ &	0.157 & $0.8^{+2.1}_{-2.1}$ & $2\pm1$ \\
RXJ$0841.1+6422$ & $7.5\pm1.3$ & $668^{+55}_{-59}$ & 	0.354 & $9.0^{+3.9}_{-3.5}$ & $15\pm4$ \\
RXJ$0847.1+3449$ & $10.3\pm1.7$ & $937^{+74}_{-80}$ &	0.452 & $24.2^{+8.9}_{-7.6}$ & $21\pm5$ \\
RXJ$0910.6+4248$ & $3.5\pm1.7$ & $551^{+121}_{-157}$ &	0.246 & $4.0^{+4.7}_{-2.9}$ & $3\pm2$ \\
RXJ$0921.2+4528$ & $2.1\pm1.5$ & $344^{+108}_{-164}$ &	0.180 & $1.2^{+1.8}_{-1.2}$ & $5\pm2$ \\
RXJ$0926.6+1242$ & $8.9\pm1.5$ & $822^{+65}_{-70}$ & 	0.407 & $16.2^{+6.0}_{-5.1}$ & $11\pm4$ \\
RXJ$0957.8+6534$ & $4.5\pm1.4$ & $605^{+89}_{-105}$ &	0.257 & $4.3^{+3.2}_{-2.6}$ & $3\pm2$ \\
RXJ$1015.1+4931$ & $6.0\pm1.4$ & $618^{+70}_{-79}$ &	0.296 & $5.6^{+3.4}_{-2.9}$ & $3\pm2$ \\
RXJ$1117.2+1744$ & $1.9\pm1.6$ & $331^{+114}_{-185}$ &	0.252 & $3.1^{+3.2}_{-2.1}$ & $6\pm3$ \\
RXJ$1117.4+0743$ & $5.5\pm1.4$ & $640^{+76}_{-86}$ & 	0.280 & $5.2^{+3.4}_{-2.8}$ & $13\pm4$ \\
RXJ$1123.1+1409$ & $5.8\pm1.4$ & $588^{+69}_{-78}$ & 	 0.271 & $4.0^{+3.8}_{-2.6}$ & $9\pm3$ \\
RXJ$1354.2-0221$ & $9.6\pm1.4$ & $895^{+64}_{-69}$ & 	 0.428 & $20.2^{+6.4}_{-5.6}$ & $23\pm5$ \\
RXJ$1524.6+0957$ & $4.3\pm1.3$ & $585^{+84}_{-98}$ & 	 0.336 & $9.5^{+4.4}_{-3.8}$ & $14\pm4$ \\
RXJ$1540.8+1445$ & $4.0\pm1.4$ & $530^{+83}_{-99}$ & 	 0.279 & $4.9^{+3.7}_{-2.5}$ & $13\pm4$ \\
RXJ$1642.6+3935$ & $2.2\pm1.5$ & $371^{+107}_{-156}$ &	 0.239 & $2.8^{+2.8}_{-1.8}$ & $5\pm2$ \\
RXJ$2059.9-4245$ & $4.4\pm1.3$ & $508^{+71}_{-82}$ & 	 0.280 & $4.4^{+3.3}_{-2.4}$ & $6\pm3$ \\
RXJ$2108.8-0516$ & $3.4\pm1.4$ & $442^{+81}_{-100}$ &	 0.210 & $1.8^{+2.2}_{-1.4}$ & $5\pm2$ \\
RXJ$2139.9-4305$ & $5.2\pm1.4$ & $572^{+73}_{-84}$ & 	 0.292 & $5.3^{+3.7}_{-2.6}$ & $9\pm3$ \\
RXJ$2146.0+0423$ & $8.5\pm1.4$ & $830^{+67}_{-73}$ & 	 0.436 & $21.0^{+6.7}_{-5.7}$ & $14\pm4$ \\
RXJ$2202.7-1902$ & $1.5\pm1.4$ & $319^{+127}_{-252}$ &  0.152 & $0.8^{+2.0}_{-0.8}$ & $2\pm1$ \\
RXJ$2328.8+1453$ & $4.9\pm1.4$ & $612^{+81}_{-93}$ &   0.254 & $4.0^{+3.7}_{-2.6}$ & $6\pm3$ \\
\hline
\hline
\end{tabular}
\tablecomments{The SIS and NFW models were fit to the data at
  $200-750h_{70}^{-1}$kpc. The listed errors include the contribution
  from large-scale structure along the line-of-sight. The listed value
  for $r_{2500}$ is derived from the best-fit NFW model. This radius
  is used to compute $N_{2500}$, the excess number of galaxies with a
  rest-frame $B$-band luminosity $-22<M_B<-18.5$ within $r_{2500}$.}
\end{center}
\end{table*}

\section{Results}

As discussed above, we consider two often used parametric models for
the cluster mass distribution. We fit these to the observed lensing
signal at $200-750h_{70}^{-1}$kpc from the cluster center.  The
resulting Einstein radii and velocity dispersions are listed in
Table~\ref{tabresult}. We also list the best-fit value for $M_{2500}$
inferred from the NFW model fit to the signal, where we use the
mass-concentration relation given by Eqn.~\ref{mcrel}. For reference
we also list the corresponding value for $r_{2500}$ from the NFW fit.

We use this radius to compute $N_{2500}$, the excess number of
galaxies with a rest-frame $B$-band luminosity $-22<M_B<-18.5$ within
$r_{2500}$ (we assume passive evolution when computing the rest-frame
$B$-band luminosity).  Recent studies of the cluster richness consider
only galaxies on the red-sequence, because of the higher contrast,
which improves the signal-to-noise ratio of the
measurement. Unfortunately we lack color information, and we compute
the total excess of galaxies.  The background count levels were
determined using the COSMOS pointings that were used to measure the
size-dependence of CTE. 

\begin{figure*}
\begin{center}
\leavevmode
\hbox{%
\epsfxsize=\hsize
\epsffile[18 160 592 710]{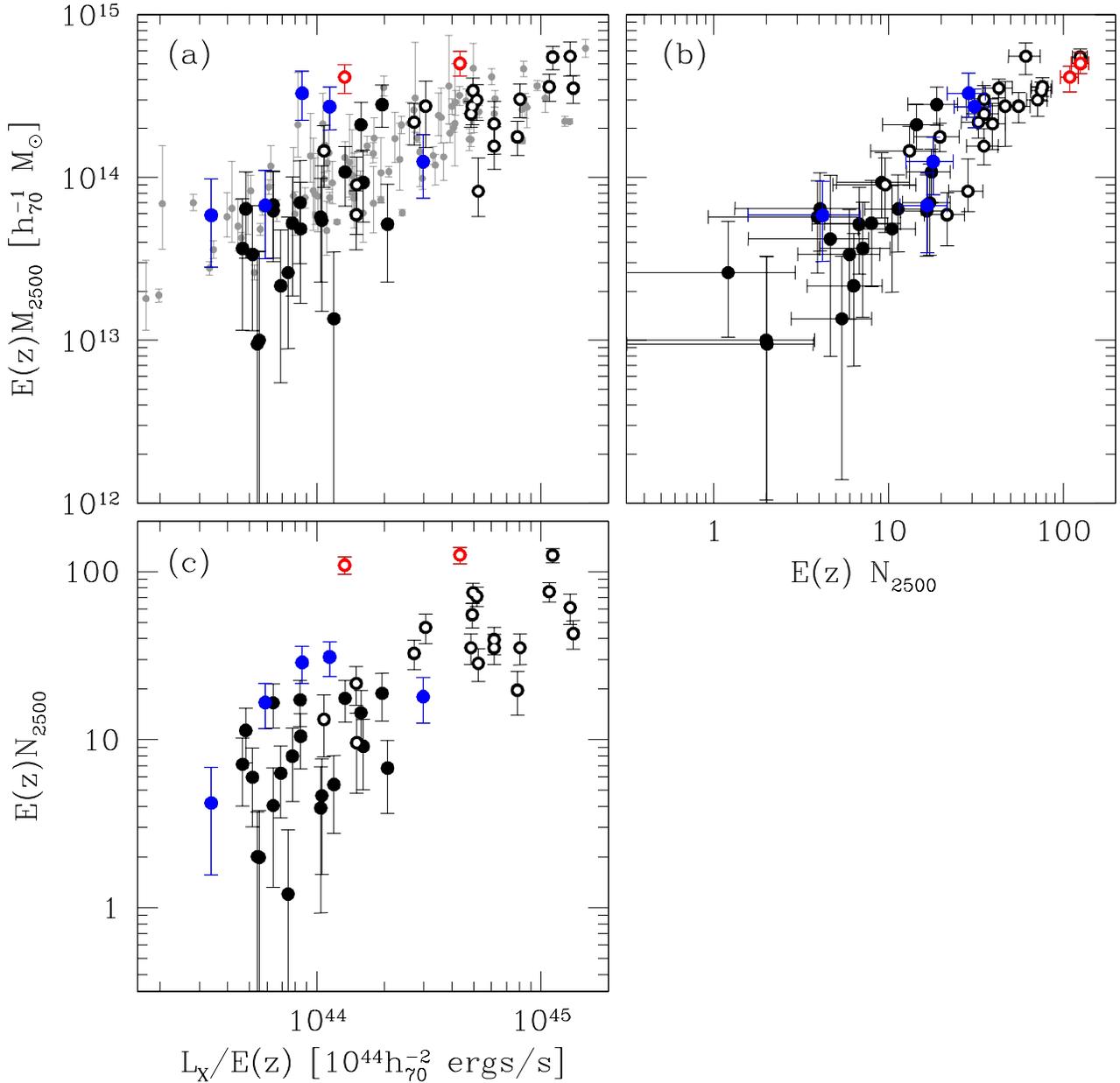}}
\caption{{\it panel a:} Plot of $M_{2500}$ as a function of X-ray
  luminosity. To account for the range in redshift of the clusters,
  the mass and luminosity have been rescaled using the corresponding
  value for $E(z)$ under the assumption of self-similar evolution. The
  solid points are the clusters from the sample studied here. The solid blue
  points are the clusters from the 160SD survey for which X-ray
  temperatures have been determined (see \S3.1). The open
  points correspond to the more massive clusters studied in
  \cite{Hoekstra07}. The open red points indicate the well-known
  strong lensing clusters A370 and CL0024+16.  For comparison the small grey points
  show the X-ray derived masses from \cite{Reiprich02}, converted to
  $M_{2500}$; {\it panel b:} Plot of $M_{2500}$ as a function of
  $N_{2500}$, the excess of galaxies with restframe $-22<M_B<-18.5$
  within $r_{2500}$; {\it panel c:} Plot of $N_{2500}$ as a function
  of X-ray luminosity. }\label{mlx}
\end{center}
\end{figure*}

Figure~\ref{mlx}a shows the resulting lensing mass as a function of
the restframe X-ray luminosity in the $0.1-2.4$ keV band. The
luminosities and masses have been scaled to redshift zero assuming
self-similar evolution with respect to the critical density 
\citep[e.g.,][]{Kaiser86,Bryan98}, where

\begin{equation}
E(z)=\frac{H(z)}{H_0}=\sqrt{\Omega_m(1+z)^3+\Omega_\Lambda}
\end{equation}

\noindent for flat cosmologies. The solid points indicate the clusters
from the sample studied here. The clusters from the 160SD survey for
which X-ray temperatures have been determined (see \S3.1) are
indicated by blue points. To extend the range in X-ray luminosity we
also show measurements for the massive clusters that were studied in
\cite{Hoekstra07}. We note, that \cite{Hoekstra07} used bolometric
X-ray luminosities from \cite{Horner01}, whereas here we use the
restframe luminosities in the $0.1-2.4$ keV band (which are a factor
$\sim 4$ smaller). We use the mass estimates from \cite{Mahdavi08}
which used new photometric redshift distributions, which were based on
much larger data sets \citep{Ilbert06}.

The agreement in lensing masses is good in the regions where the two
samples overlap. However, the scatter in the mass-luminosity relation
is substantial (both for the clusters studied here as well as the more
massive clusters). We examined whether some of the scatter could be
due to the uncertainty in the position of the cluster center, but find
no difference when comparing results for clusters with different
levels of confidence in the identification of the BCG (see $Q_{\rm
  BCG}$ in Table~\ref{tabsample}. It is worth noting, however, that we
do not find any massive clusters $(>10^{14}M_{\odot})$ that are X-ray
faint (i.e., $L_X<10^{44}$~erg/s), implying that the dispersion in the
$M-L_X$ relation is relatively well-behaved.

\cite{Reiprich02} studied a sample of 63 X-ray bright clusters and derived
masses under the assumption of hydrostatic equilibrium. This sample of
clusters spans a similar range in $L_X$ as our combined sample. We
converted their measurements for $M_{500}$ to $M_{2500}$ using the
mass-concentration relation given by Eqn.~\ref{mcrel} and show the
results in Figure~\ref{mlx} (small grey points). The agreement with
our findings is very good. A more quantitative comparison is presented
in \S3.3.

\cite{Yee03} have shown that there is a good relation between the
cluster richness and the mass (and various proxies) for massive
clusters. Our study extends to lower masses and as is shown in
Figures~\ref{mlx}b and~c $N_{2500}$ correlates well with both the
X-ray luminosity and lensing mass\footnote{To compute $N_{2500}$ for
  the sample of massive clusters we used the same selection criteria
  as discussed above.}. Similar results have been obtained using SDSS
cluster samples \citep[e.g.,][]{Rykoff08a,Johnston07}.  We assumed
that $N_{2500}$ scales as the mass $M_{2500}$, which is a reasonable
assumption if the galaxies trace the density profile. This choice,
however, does not affect our conclusions.

The results agree well in the regions where the two samples overlap.
The correlation between $N_{2500}$ and $M_{2500}$ is tighter than that
of $N_{2500}$ and $L_X$. The former is less sensitive to the
projections along the line-of-sight (either substructures or an
overall elongation of the cluster), because both the galaxy counts and
the lensing mass are derived from projected measurements.  The X-ray
results provide a different probe of the distribution of baryons,
which is expected to lead to additional scatter.  Furthermore, some of
the scatter may be caused by unknown contributions by AGNs. The
importance of AGN can be evaluated using a combination of deeper, high
spatial resolution $(\lesssim 5'')$ X-ray and radio imaging. Such X-ray
observations would also provide estimates for the temperature of the
X-ray gas (which is a better measure of the cluster
mass). Unfortunately, such data exist for only five of the low-mass
clusters. For these, X-ray temperatures have been derived, which are
listed Table~\ref{tabtx}.  For the massive clusters we use the values
from \cite{Horner01} that were used in \cite{Hoekstra07}. We note that
all clusters follow a tight $L_x-T_x$ relation.

\subsection{Comparison with X-ray temperature}

\begin{figure}
\begin{center}
\leavevmode
\hbox{%
\epsfxsize=8.5cm
\epsffile{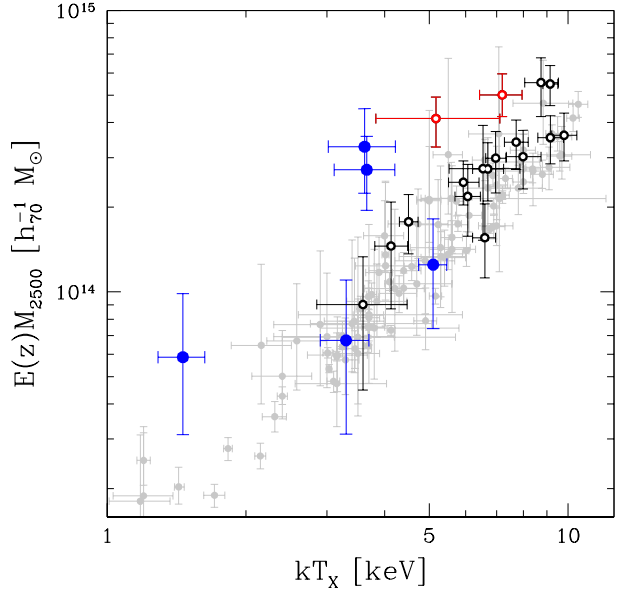}}
\caption{Plot of $M_{2500}$ as a function of X-ray temperature. To
  account for the range in redshift of the clusters, the mass has been
  rescaled using the corresponding value for $E(z)$ under the
  assumption of self-similar evolution. The solid blue points are the
  clusters from the sample studied here. The open points correspond to
  the more massive clusters studied in H07, whereas the small grey
  points show the results from Reiprich \& B{\"o}hringer
  (2002). The well-known strong lenses A370 and CL0024+16 are indicated
  as red open points.}\label{mtx}
\end{center}
\end{figure}

Figure~\ref{mtx} shows the resulting plot of $M_{2500}$ as a function
of X-ray temperature. RXJ$1117.4+0743$ and RXJ$1524.6+0957$ lie on the
tight relation defined by the bulk of the clusters. The measurements
from \cite{Reiprich02} also follow this relation (light grey
points). However, RXJ$0847.1+3449$ and RXJ$1354.2-0221$ appear to be
more massive than might be expected based on $T_X$. They appear to lie
on a parallel relation, along with some of the clusters from
\cite{Hoekstra07}. The latter clusters are CL0024+16 and A370, which
are well known strong lensing clusters (indicated in red in
Figures~\ref{mlx} and~\ref{mtx}). These clusters were observed because of their strong
lensing properties and included in the search for archival CFHT data
carried out by \cite{Hoekstra07}.

Interestingly, all four clusters are outliers on both the
$M_{2500}$-$L_x$ and $N_{2500}-L_x$ relations presented in
Figure~\ref{mlx}, but follow the mass-richness relation. This is
consistent with the presence of (sub)structures along the
line-of-sight boosting both $M_{2500}$ and $N_{2500}$: the projection
of two mass concentrations (along the line-of-sight) would increase
the richness and the weak lensing mass approximately linearly.  The
inferred X-ray temperature on the other hand will be close to that of
the more massive system, whereas the X-ray luminosity will be much
lower than expected, because it is proportional to the square of the
electron density. We note that both RXJ$0847.1+3449$ and
RXJ$1354.2-0221$ show evidence of strong lensing.

\begin{table}
\begin{center}
\caption{X-ray temperatures
\label{tabtx}}
\begin{tabular}{lll}
\hline
\hline
name\hspace{2.5cm} & $kT_X$ [keV]\hspace{1cm} & ref. \\
\hline
RXJ$0110.3+1938$ & $1.46^{+0.19}_{-0.15}$ & 1 \\
RXJ$0847.1+3449$ & $3.62^{+0.58}_{-0.51}$ & 2 \\
RXJ$1117.4+0743$ & $3.3^{+0.42}_{-0.36}$  & 3 \\
RXJ$1354.2-0221$ & $3.66^{+0.6}_{-0.5}$   & 2 \\ 
RXJ$1524.6+0957$ & $5.1\pm0.36$         & 4 \\
\hline
\hline
\end{tabular}
\tablecomments{references: (1) \cite{Bruch10}; (2) \cite{Lumb04};
  (3) \cite{Carrasco07}; \cite{Vikhlinin02}}
\end{center}
\end{table}

Interestingly, the X-ray image of RXJ$0847.1+3449$ in \cite{Lumb04}
shows evidence for a nearby cluster candidate. The case is less clear
for RXJ$1354.2-0221$, but the X-ray image shows a complex
morphology. Unfortunately we lack the dynamical data to confirm
whether RXJ$0847.1+3449$ and RXJ$1354.2-0221$ are projected systems.
The cluster RXJ$1117.4+0743$ was studied in detail by
\cite{Carrasco07}, who find that this cluster is also a projection of
two structures. For the main component \cite{Carrasco07} infer a
galaxy velocity dispersion of $592\pm82$ km/s, whereas the second
structure is less massive with $\sigma_v=391\pm85$ km/s. Based on our
lensing analysis we obtained a velocity dispersion of
$639^{+76}_{-86}$ km/s, in good agreement with the dynamical results
for the main cluster. \cite{Carrasco07} also performed a weak lensing
analysis based on ground based imaging data and obtained a velocity
dispersion of $\sigma=778\pm89$ km/s (where we took the average of
their results for the $g'$ and $r'$ band), implying a mass 50\% larger
than our estimate. We are not able identify an obvious cause for this
difference, but note that PSF-related systematics have a larger impact
on ground based results.

\subsection{The $M_{2500}-L_X$ scaling relation}

In this section we examine the correlation between $M_{2500}$ and the
X-ray luminosity, in particular the normalization and the power law
slope. We fit a power law model to the combined sample to maximize the
leverage in X-ray luminosity

\begin{equation}
E(z)M_{2500}=M_x \left(\frac{L_x/E(z)}{2 \times 10^{44}{\rm erg/s}}\right)^\alpha.
\end{equation}

If we naively fit this model to the measurements we find that the
value for $\chi^2$ of the best fit is too high ($\chi^2=106$ with 43
degrees of freedom). This indicates that there is intrinsic scatter in
the relation, which is also apparent from Figure~\ref{mlx}.  We need
to account for the intrinsic scatter in the fitting procedure, because
ignoring it will generally bias the best fit parameters.  We fit the
model to our measurements, which have errors that follow a normal
distribution. The intrinsic scatter, however, can be described by a
log-normal distribution \citep[see e.g., Fig. 13
  in][]{Vikhlinin09a}. We will assume that the intrinsic scatter can
be approximated with a normal distribution with a dispersion
$\sigma_Q\approx \ln(10)Q\log Q$ (we use the log with base 10).

To fit the $M_{2500}-L_X$ relation we follow a maximum likelihood
approach. For a model $f$ with parameters ${\bf a}$, the predicted
values are $y_i=f(x_i;{\bf a})$. The uncertainties in $x_i$ and $y_i$
are given by $\sigma_{x,i}$ and $\sigma_{y,i}$. If we assume a
Gaussian intrinsic scatter $\sigma_{Q,i}$ in the $y_i$ coordinate, the
likelihood ${\cal L}$ is given by

\begin{equation}
{\cal L}=\prod_{i=1}^{n}\frac{1}{\sqrt{2\pi}w_i}\exp\left[-\frac{[y_i-f(x_i;{\bf a})]^2}{2w_i^2}\right],
\end{equation}

\noindent where $w_i$ accounts for the scatter:

\begin{equation}
w_i^2=\left[{\frac{df}{dx}(x_i)}\right]^2\sigma_{x,i}^2+\sigma_{y,i}^2+\sigma_{Q,i}^2.
\end{equation}

\noindent If we consider the logarithm of the likelihood it becomes clear
why including the intrinsic scatter differs from standard least squares
minimization:

\begin{equation}
-2\ln {\cal L}=2\sum_{i=1}^n \ln w_i + \sum_{i=1}^n \left(\frac{y_i-f(x_i;a_j)}{w_i}\right)^2+C,
\end{equation}

\noindent where the second term corresponds to the usual $\chi^2$. For
a linear relation with no intrinsic scatter, the first term is a
constant for a given data set and the likelihood is maximimized by
minimizing $\chi^2$. However, if intrinsic scatter is included as a
free parameter, the first term acts as a penalty function, and cannot
be ignored.

The presence of intrinsic scatter also exacerbates the Malmquist bias
for a flux limited sample, such as the 160SD
survey\footnote{We assume that the clusters studied in
  \cite{Hoekstra07} do not suffer from Malmquist bias.}.  As a result
the average flux of the observed sample of clusters is biased high
compared to the mean of the parent population (in particular near the
flux limit of the survey). To account
for Malmquist bias we follow the procedure outlined in Appendix A.2 of
\cite{Vikhlinin09a} and correct the X-ray luminosities before fitting
the $M_{2500}-L_X$ relation. We find that the correction for Malmquist
bias is relatively modest, increasing the normalisation $M_X$ by $\sim
5\%$ and reducing the slope $\alpha$ by $\sim 5\%$\footnote{We also
  computed the cluster mass function in a $\Lambda$CDM cosmology
  (using parameters from \cite{Evrard02} and assigned X-ray
  luminosities using our best fit $M_{2500}-L_X$ relation and
  intrinsic scatter and find similar biases.}.

We find an intrinsic scatter of $\sigma_{\log
  L_X|M}=0.23^{+0.10}_{-0.04}$ (or a relative error of $\sim 70\%$),
in good agreement with other studies. \cite{Reiprich02} list a
somewhat larger scatter of $\sigma_{\log L_X|M}=0.29$ for the HIFLUGCS
sample, whereas \cite{Vikhlinin09a} and \cite{Mantz10b} find
$\sigma_{\log L_X|M}=0.17\pm0.02$ and $0.18\pm0.02$, respectively. Our
results also agree well with \cite{Stanek06} who find $\sigma_{\log
  M|L_X}=0.19\pm0.03$, which is in good agreement with our value of
$0.17^{+0.04}_{-0.03}$.

\begin{figure}
\begin{center}
\leavevmode 
\hbox{%
\epsfxsize=8.5cm 
\epsffile{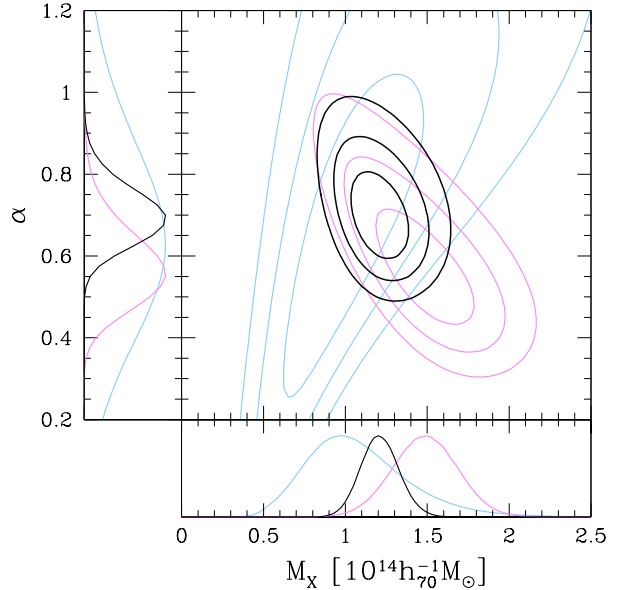}}
\caption{Likelihood contours for the slope of the $M_{2500}-L_X$
  relation and the mass of a cluster with a luminosity of $L_X=2\times 10^{44}
  h_{70}^{-2}$ ergs/s. The contours indicate the 68.3\%, 95.4\% and
  99.7\% confidence limits on two parameters jointly. The side panels
  show the probability density distributions for each parameter (while
  marginalising over the other). The black curves correspond to the
  constraints determined from the combined sample of clusters, whereas
  the cyan and pink curves correspond to the low mass and high
  mass samples, respectively. 
}\label{slope}
\end{center}
\end{figure}

The likelihood contours for the power law slope $\alpha$ of the
$M_{2500}-L_X$ relation and the mass $M_{2500}$ of a cluster with a
luminosity of $L_X=2\times 10^{44} h_{70}^{-2}$erg/s are shown in
Figure~\ref{slope}.  For the combined sample of clusters we find a
best fit slope of $\alpha=0.68\pm0.07$ and a normalization
$M_X=(1.2\pm0.12)\times 10^{14} h_{70}^{-1}\msun$. The inferred
slope is consistent with the value of $\frac{3}{4}$ expected for self-similar
evolution \citep{Kaiser86}.

It is also interesting to compare the constraints from both
sub-samples separately. The best fit slope for the clusters studied in
\cite{Hoekstra07} is $\alpha=0.55^{+0.10}_{-0.09}$ with a
normalization of $M_X=(1.5\pm0.2)\times 10^{14}h_{70}^{-1}\msun$
(indicated by the pink contours)\footnote{We note that the slope is
  different from the original \cite{Hoekstra07} results, who
  found $\alpha=0.43\pm0.1$. The reason for this large change is
  twofold.  First, \cite{Hoekstra07} did not account for intrinsic
  scatter in the fit of the scaling relation. Furthermore, the current
  analysis includes three clusters for which no X-ray data was
  available in \cite{Hoekstra07}.  The clusters in question are A209,
  A383 and MS1231+15. The latter two have the lowest X-ray
  luminosities and drive much of the change in slope, whereas
  including A209 does not change the previous results. Restricting the
  sample to the clusters that were used by \cite{Hoekstra07} to
  constrain the slope, we find $\alpha=0.41\pm0.10$, in agreement with
  the orginal result. Such a rather large variation in best-fit slope
  demonstrates the need for larger samples of clusters with
  multi-wavelength data.}.  The blue contours in Figure~\ref{slope}
indicate the constraints for the HST sample studied here. The
parameters are not well constrained, with best fit values
$\alpha=0.63\pm 0.24$ and a normalization
$M_X=(1.0\pm0.24\times 10^{14}h_{70}^{-1}\msun$.  

The difference in the constraints from the (extended)
\cite{Hoekstra07} sample and the 160SD systems may hint at a deviation
from a single power law relation. As discussed in \S2.5, however, the
uncertainty in the position of the cluster center leads to an
underestimate of the cluster mass, as does the presence of
substructure. The \cite{Hoekstra07} results are much less sensitive to
these problems. To quantify this, we combine the \cite{Hoekstra07}
results with the 160SD clusters with $Q_{\rm BCG}=2$ (12 systems) and
the ones with $Q_{\rm BCG}<2$ (13 systems). For the former we find
$M_X=(1.30\pm0.15)\times 10^{14}h_{70}^{-1}\msun$ and
$\alpha=0.63\pm0.08$, whereas requiring $Q_{\rm BCG}<2$ yields
$M_X=(1.19\pm0.14)\times 10^{14}h_{70}^{-1}\msun$ and
$\alpha=0.68\pm0.08$. This comparison suggests that our normalisation
may be biased low by as much as $\sim 10\%$.

\subsection{Comparison with X-ray samples}

\begin{figure}
\begin{center}
\leavevmode 
\hbox{%
\epsfxsize=8.5cm 
\epsffile{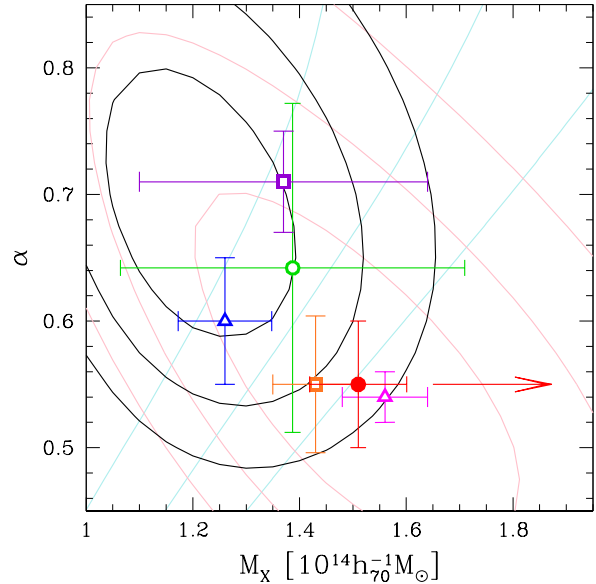}}
\caption{Comparison of the slope of the $M_{2500}-L_X$ relation and
  the mass of a cluster with a luminosity of $L_X=2\times 10^{44}
  h_{70}^{-2}$ ergs/s. The contours are the same as in
  Figure~\ref{slope}. The blue triangle indicates the results from
  \cite{Reiprich02}. The result from \cite{Stanek06} is
  indicated by the open pink triangle. We also show constraints from
  two recent studies of massive clusters by \cite{Vikhlinin09a}
  (orange square) and \cite{Mantz10b} (purple square). The results
  from a study of X-ray groups in COSMOS by \cite{Leauthaud10} are
  indicated by the green circle. Note that the accuracy of the
  comparison is limited, because of differences in the
  mass-concentration relation. The results from \cite{Rykoff08b} based
  on SDSS data are indicated by the red point. The arrow indicates the
  possible shift in mass because of a bias in the source redshift
  distribution.}\label{zoom}
\end{center}
\end{figure}

The relation between the X-ray luminosity and cluster mass has been
studied extensively. In this section we compare our measurements to a
number of recent results, which are shown in Figure~\ref{zoom}. Where
needed, the X-ray luminosities are converted to the $0.1-2.4$ keV band
and Eqn.~\ref{mcrel} has been used to convert masses to $M_{2500}$ and
adjust the slopes (because the relation between $M_{2500}$ and
$M_{200}$ (or $M_{500}$) is a power law with a slope less than 1.).

\cite{Reiprich02} studied a sample of 63 clusters, with masses derived
under the assumption of hydrostatic equilibrium. Their BCES bisector
results for the flux-limited sample yields $\alpha=0.60\pm0.05$ and
$M_x=(1.3\pm0.09)\times10^{14} \msun$ are indicated by the blue
triangle in Fig.~\ref{zoom}. \cite{Stanek06}, however, have argued
that these results suffer from Malmquist bias. Instead they compared
the X-ray number counts to the mass function in $\Lambda$CDM
cosmologies and derived $\alpha=0.54\pm0.02$ and a high normalization
of $M_X=(2.1\pm0.1)\times 10^{14}\msun$. This result, however depends
strongly on the adopted value for $\sigma_8$, and combination with the
WMAP3 data \citep{Spergel07} lowers the normalization to
$M_X=(1.56\pm0.08)\times10^{14}\msun$ (open pink triangle in
Fig.~\ref{zoom}).  \cite{Vikhlinin09a} studied a sample of $z\sim
0.05$ and $z\sim 0.5$ clusters that were observed with Chandra. Their
results with $\alpha=0.55\pm0.05$ and
$M_x=(1.43\pm0.08)\times10^{14}\msun$ are indicated by the open orange
square. Another recent study was presented by \cite{Mantz10b} who
found a fairly steep slope of $\alpha=0.70\pm0.04$ and
$M_x=(1.42\pm0.28)\times10^{14}\msun$ (purple square).

The constraints from these studies are largely driven by X-ray
luminous clusters (this is particularly true for the results of
\cite{Vikhlinin09a} and \cite{Mantz10b}). If the slope of the
$M_{2500}-L_X$ varies with $L_X$, then it is more appropriate to
compare these studies to the results from the sample studied in
\cite{Hoekstra07}, for which the agreement is indeed very good.  When
combined with the results from the 160SD survey, the resulting scaling
relation has a lower normalisation and steeper slope. To examine
whether this could be caused by a change in slope, it is interesting
to compare to measurements at lower luminosities.

At the low X-ray luminosity end, \cite{Leauthaud10} studied X-ray
groups using COSMOS. Because of the small area surveyed, the groups
are less luminous than the sample of clusters from the 160SD survey
studied here. Nonetheless it is interesting to extrapolate their
results for comparison. \cite{Leauthaud10} use the mass-concentration
relation from \cite{Zhao09}. We refit the 160SD sample using this
relation and find that the average $M_{2500}$ is unchanged, but that
$M_{200}$ is increased by 24\%. We account for this when converting
the measurements from \cite{Leauthaud10}, and find that they imply
$M_x=(1.4\pm0.3)\times10^{14}\msun$ and a slope $\alpha=0.61\pm0.13$
(indicated by the open green circle in Fig.~\ref{zoom}), in fair
agreement with our results. We also note that this highlights the
difficulty in comparing results, especially when the analyses differ
in detail. This is particularly relevant for the comparison with SDSS
results in the following section.

\subsection{Comparison with SDSS}

\cite{Rykoff08b} measured the scaling relation between $L_X$ and
$M_{200}$ for a large sample of optically selected clusters found in
the Sloan Digital Sky Survey \citep{Koester07}. This allowed them to
extend the study of the $M-L_X$ relation to much lower luminosities,
compared to most of the measurements presented in the previous
section.  The clusters were binned in richness and for each bin, the
mean X-ray luminosity and weak lensing mass were determined.  In the
lensing analysis, described in \citep{Johnston07}, both the mass and
concentration are fit as free parameters. The resulting values for $c$
agree well with the relation presented in \cite{Duffy08}. Upon
conversion to $M_{2500}$ we find that the measurements from
\cite{Rykoff08b} correspond to $M_X=(1.5\pm0.1)\times
10^{14}h_{70}^{-1}\msun$ and $\alpha=0.55\pm0.04$. This result is
indicated in Figure~\ref{zoom} by the red point.

\cite{Mandelbaum08b} have pointed out that the weak lensing masses
determined by \cite{Johnston07} may be too low (by as much as $\sim
24\%$), because of a bias in the source redshift distribution
\citep[also see][]{Leauthaud10}. The red arrow indicates the shift in
normalization if the bias pointed out by \cite{Mandelbaum08b} is
correct. In that case the results from \cite{Rykoff08b} disagree with
all other measurements. 

\cite{Rykoff08b} find a shallower slope and higher mass, which appears
to be inconsistent with our results. As discussed above, uncertainties
in the adopted cluster centers may lead us to underestimate our
normalisation by as much as $\sim 10\%$, which is not sufficient to
remove the difference. \cite{Johnston07} model the centroiding
uncertainty using results from mock catalogs. However, if the model
overestimates the offsets the resulting masses will be biased high and
\cite{Mandelbaum08a} argue that this may indeed be the case.

Another possible explanation for the large normalisation of
\cite{Rykoff08b} is that their results are expected to be biased
towards a lower $L_X$ at a given lensing mass. The reason for this was
already alluded to in \S3.1, namely that some clusters appear X-ray
underluminous and have low X-ray temperatures, given their high
masses. These clusters, however do follow the tight mass-richness
relation. \cite{Rykoff08b} bin their clusters in richness and analyse
stacked X-ray and lensing data. The fact that $N_{2500}$ and
$M_{2500}$ are strongly correlated, then leads to a larger mean mass
at a given X-ray luminosity.

The presence of substructure in the cluster (as well as filaments),
will boost both the lensing and richness estimates relative to the
X-ray luminosity. Such structures are numerous at the low mass end: it
is rare to find an alignment of rich clusters, because they are rare,
whereas an alignment of groups should be more frequent. Hence, the
bias is likely to be mass dependent, increasing the masses of low
X-ray luminosity systems by a larger fraction, compared to X-ray
luminous clusters. The consequence is a flatter slope of the $M-L_X$
relation.

Comparison with numerical simulations can help clarify the amplitudes
of the various biases listed above. However, the argument presented
above highlights the danger of stacking samples of clusters when
parameters are correlated. Identifying and quantifying such
correlation requires well-characterized (both in selection and data
coverage) and large samples of clusters. The 160SD clusters provide an
excellent starting point, but increasing the number of cluster in this
mass range with high quality X-ray data is of great importance to
better constrain the scaling relations and to help interpret the
results from stacked samples of clusters.

\section{Conclusions}

To extend the observed scaling relation between cluster mass and X-ray
luminosity towards lower $L_X$, we have determined the masses of a
sample of 25 clusters of galaxies drawn from the 160 square degree
ROSAT survey \citep{Vikhlinin98}. The clusters have redshifts
$0.3<z<0.6$, and the X-ray luminosities range from $2\times 10^{43}$
to $2\times 10^{44}$ erg/s. The masses were determined based on a weak
lensing analysis of images in the $F814W$ filter obtained using the
ACS on the HST. 

To measure the mass we assume that the brightest cluster galaxy
indicates the center of the cluster. In most cases this leads to an
unambiguous identification, but in a number of cases the choice of
center is less clear. Nonetheless, we have verified that our choice of
center does not affect the results for the inferred scaling relations.

To correctly interpret the weak lensing data, we derived an accurate
emperical correction for the effects of CTE on the shapes of faint
galaxies. We detect a significant lensing signal around most of the
clusters. To increase the range in cluster properties, we extend the
sample with massive clusters studied by \cite{Hoekstra07}. The lensing
masses agree well where the two samples overlap in $L_X$.

The inferred lensing masses correlate well with the overdensity of
galaxies (i.e., cluster richness). The relation between mass and X-ray
luminosity has significant intrinsic scatter. Under the assumption it
follows a log-normal distribution we find a scatter of $\sigma_{\log
  L_X|M}=0.23^{+0.10}_{-0.04}$, which is in good agreement with other
studies. We fit a power law relation between $M_{2500}$ and $L_X$ and
find a best fit slope of $\alpha=0.68\pm0.07$ and a normalisation (for
$L_X=2\times 10^{44} h_{70}^{-2}$ erg/s) of $M_X=(1.2\pm0.12)\times
h_{70}^{-1} 10^{14}\msun$. Comparison with other studies is
complicated by the fact that the conversion of the masses depends on
the assumed mass-concentration relation. We find that the results for
the sample of massive clusters from \cite{Hoekstra07} agrees well with
a number of recent studies. The combination with clusters from the
160SD survey lowers the normalisation, which could be caused by a
steepening of the $M-L_X$ relation. However, a study of low mass
systems by \cite{Rykoff08b} finds a higher normalisation.

Uncertainties in the position of the cluster center, as well as
deviations from the adopted NFW profile (e.g., substructures) may bias
our masses low, but we estimate this to be less than 10\%, which
cannot explain the difference with \cite{Rykoff08b}. On the other
hand, structures along the line-of-sight, which also may simply
reflect the fact that clusters themselves are highly elongated, will
lead to a higher normalisation for the study by \cite{Rykoff08b}.
They binned clusters discovered in the SDSS by richness and measured their
ensemble averaged X-ray luminosity and lensing mass. In this case the
tight correlation between lensing mass and richness results in a low
average $L_X$ at a given mass. Furthermore, the relative importance of
substructures and projections is mass dependent, preferentially
affecting low mass systems. Consequently, both the slope and the
normalization of the $M-L_X$ relation are affected.

To investigate the importance of the various biases, larger samples
of low mass clusters need to be observed. Better X-ray observations
provide a good starting point to extend the mass range over which
scaling relations are determined and to improve the interpretation
of ensemble averaged samples of clusters.

\acknowledgments 

We thank the anonymous referee for useful suggestions that have
improved the paper. HH acknowledges support by NSERC, the Canadian
Institute for Advanced Research (CIfAR) and the Canadian Foundation
for Innovation (CFI) as well as a VIDI grant from the Nederlandse
Organisatie voor Wetenschappelijk Onderzoek (NWO). HH also thanks
Andisheh Mahdavi for discussion on fitting relations with intrinsic
scatter. MD and GMV acknowledge the support of two STScI/HST grants
(HST-GO-10490.01 HST-GO-10152.01) and and a NASA LTSA grant
(NNG-05GD82G).  This research has made use of the X-Rays Clusters
Database (BAX) which is operated by the Laboratoire d'Astrophysique de
Tarbes-Toulouse (LATT), under contract with the Centre National
d'Etudes Spatiales (CNES).

\bibliography{snap}

\appendix

\section{Measurement of CTE effects}

In this appendix we present the results of our study of the effect of
CTE on shape measurements. Similar to other studies
\citep[e.g.,][]{Massey07,Rhodes07} we derive an empirical correction,
although our actual implementation differs from these previous works
in a number of ways. 

The CTE problem occurs during readout, and therefore the correction
should be applied before the correction for PSF anisotropy. Ideally
one would like to correct the images before shape measurements are
done \citep{Massey10}, but for our purposes such a sophisticated
approach is less important. Instead we will quantify the change in
$e_1$ (this is the polarization component that quantifies the
elongation along the $x-$ and $y-$ direction). We note that our
measurements are done on images after they have been corrected for
camera distortion.

\cite{Massey07} measured the effect of CTE by determining the
mean galaxy ellipticity as a function of distance from the readout
electronics. We cannot do so for a number of reasons. First, we have
data from a much smaller number of images. More importantly, the
presence of the lensing signal induced by the clusters also gives rise
to a variation in $e_1$: it is larger in the centre of the field and
decreases towards the edge. Instead we examine observations of dense star
fields. After correction for PSF anisotropy correction, the shape of
each star provides an accurate estimate of the CTE effect, because it
is intrinsically round, unlike a galaxy. As a result, we can measure
the amplitude of CTE much more accurately.

\begin{figure*}
\begin{center}
\leavevmode
\hbox{%
\epsfxsize=8.5cm
\epsffile{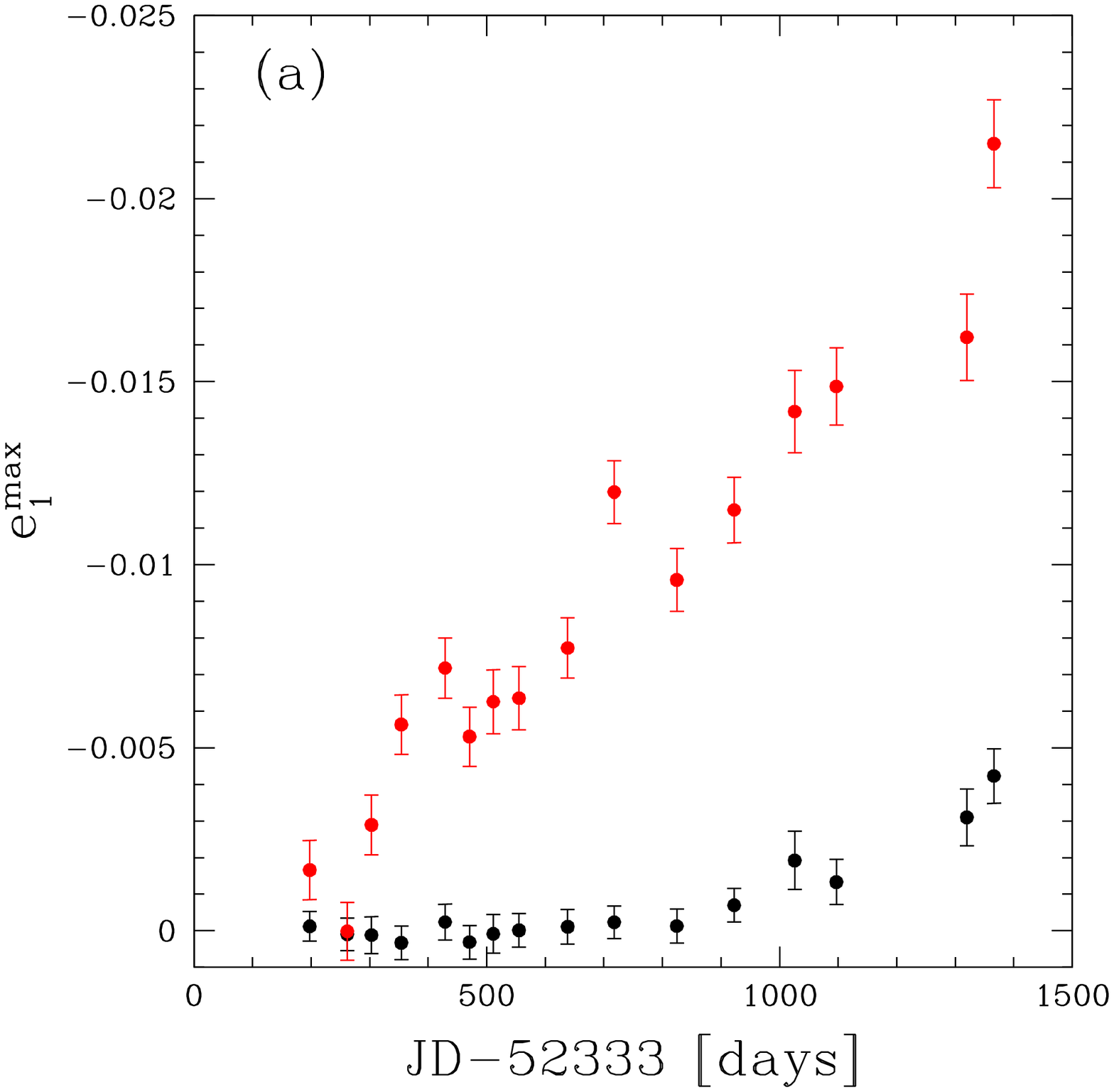}
\epsfxsize=8.5cm
\epsffile{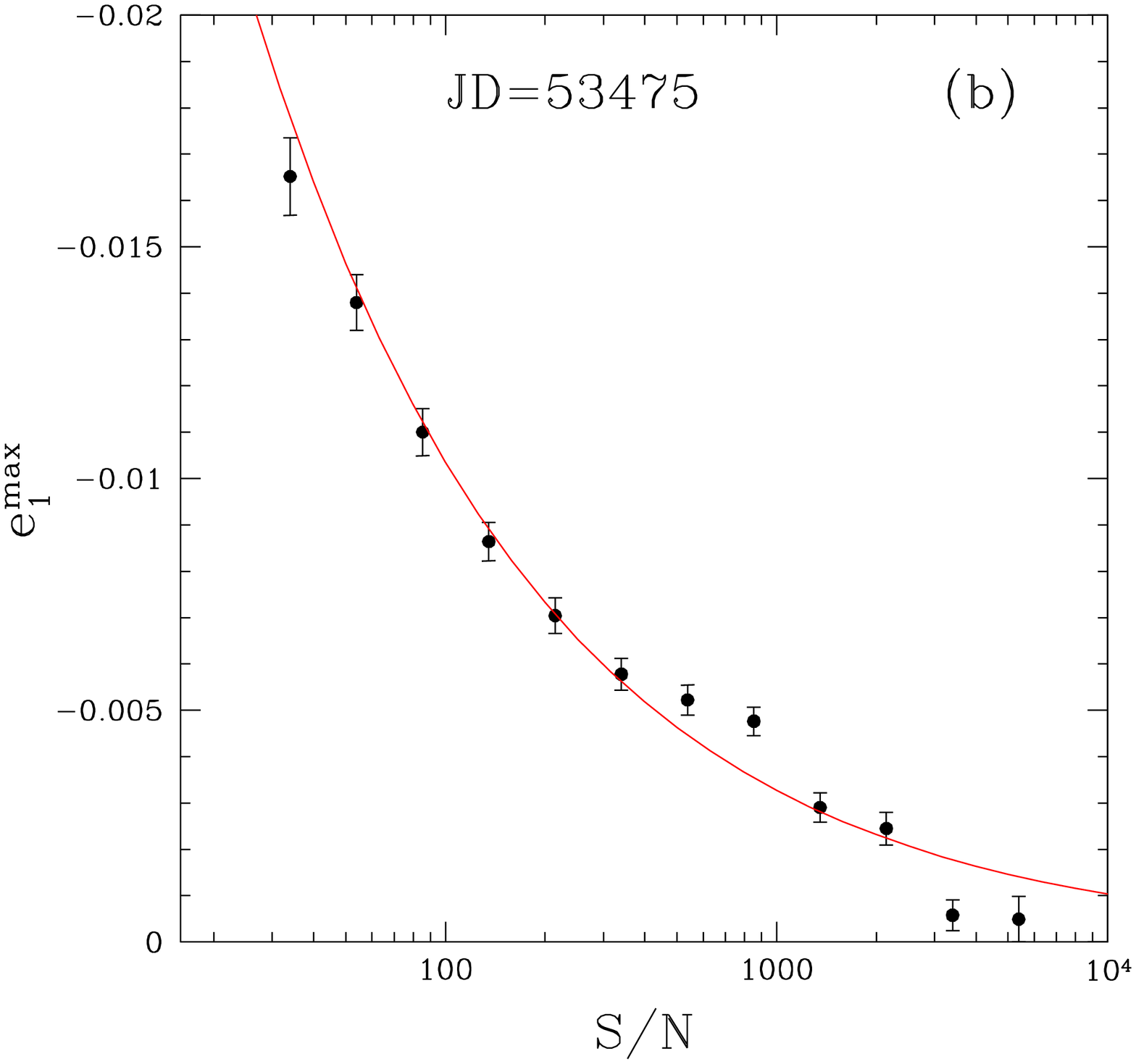}}
\caption{{\it Panel a:} CTE induced polarization at the center of the
exposure (maximum distance from the read-out electronics), as a function
of time. The black points correspond to stars with a signal-to-noise
ratio of $\sim 4000$, whereas the red points show the trend for
$S/N\sim 40$. The results are consistent with a linear increase with
time. {\it Panel b:} maximum CTE induced polarization
as a function of S/N ratio at the mean Modified Julian Date of our cluster
observations. The red curve is the best fit model $\propto \sqrt{S/N}$.
Note that this scaling is different from \cite{Rhodes07}.
\label{cte_star}}
\end{center}
\end{figure*}

We use observations of the star cluster NGC104, which has been
observed at regular epochs for a program to study the stability of ACS
photometry. We select only the data with exposure times of 30s, which
yields a uniform data set of 17 exposures, of which we omit one. These
single exposures are drizzled and we measure the shape parameters from
these images.  The PSF anisotropy model for each exposure is
determined using stars with $F814W<18$, as the effects of CTE are
expected to be small for such bright objects. The fainter stars are
corrected for PSF anisotropy and we measure $e_1$ as a function of $y$
coordinate. 

We assume that the pattern is the same for each ACS chip and combine
the data. We note that this assumption is actually supported by our
measurements. For each exposure we fit a linear trend with distance
from the read-out electronics

$$ e_1^{\rm CTE}=e_1^{\rm max} (y/2048),$$

\noindent where $e_1^{\rm max}$ is the the maximum induced
polarization. Figure~\ref{cte_star}(a) shows the resulting value for
$e_1^{\rm max}$ as a function of time for stars with high (black
points) and low (red points) signal-to-noise ratios. These
measurements are consistent with a linear increase with time of the
CTE induced distortion.

As expected the CTE effects are also more pronounced for fainter
objects. To investigate this trend further, we compute $e_1^{\rm max}$
as a function of signal-to-noise ratio, adopting the average Modified
Julian Date of our cluster observations. The results are shown in
Figure~\ref{cte_star}(b). We fit the following model to our measurements

\begin{equation}
e_1^{\rm CTE}=e_1^{\rm max} \frac{1}{\sqrt{S/N}}\left(\frac{y}{2048}\right)
(MJD-52,333).
\end{equation}

The red line in Figure~\ref{cte_star}(b) corresponds to the best
model, for which we find $e_1^{\rm max}=(-9.07\pm0.16)\times
10^{-5}$. Note that we have assumed that the CTE effect is
proportional to $\sqrt{S/N}$, which is different from \cite{Rhodes07}
who argue for a scaling $\propto (S/N)^{-1}$. The latter scaling,
however, is inconsistent with our findings. We have confirmed our
results with other stellar fields. We note that \cite{Rhodes07} use a
different shape measurement method, which may explain some of the
differences.

We now have an accurate model for the effects of CTE for stellar
images, but it is not clear whether it is adequate for galaxies, which
are more extended. Our cluster data cannot be used for this test,
because the lensing signal induced by the clusters leads to a
comparable change in $e_1$ with $y$ position. Instead we retrieved 100
pointings taken as part of the COSMOS survey (PID:10092) and analysed
these data in a similar fashion as our own.

\begin{figure}
\begin{center}
\leavevmode
\hbox{%
\epsfxsize=8.5cm
\epsffile{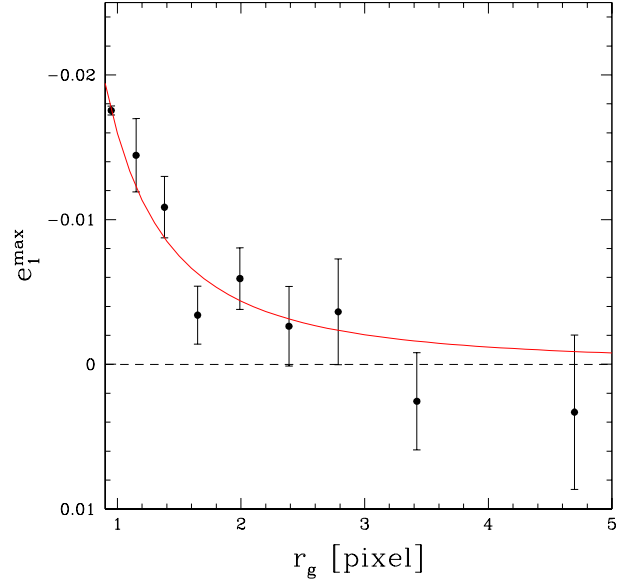}}
\caption{Maximum CTE induced polarization as a function of galaxy size $r_g$
(for a S/N=20 and MJD=53195) based on the analysis of 100 pointings of
the COSMOS survey. The left most point corresponds to the
results from our analysis of star fields. We detect a clear dependence
on size. The best fit power-law model is indicated by the solid line
and has a slope $\sim -2$.
\label{cte_gal}}
\end{center}
\end{figure}

Figure~\ref{cte_gal} shows the measurement of $e_1^{\rm max}$ as a
function of galaxy size $r_g$ for a signal-to-noise ratio of 20 and a
mean modified Julian date of 53195. The points have been corrected for
the small increase in mean S/N ratio as $r_g$ increases. The left-most
point is the result derived from NGC104. We detect a clear size
dependence of the CTE signal. We assume the dependence is a power-law,
and we find a best fit slope of $-1.85\pm0.3$; the best fit model is
indicated by the solid line in Figure~\ref{cte_gal}.  We therefore
revise our PSF-based model to account for the size dependence of the
CTE effect:

\begin{equation}
e_1^{\rm CTE}=e_1^{\rm max} \frac{1}{\sqrt{S/N}}\left(\frac{y}{2048}\right)
\left(\frac{r_g}{0.05''}\right)^{-1.85} (MJD-52,333),
\end{equation}
where the best fit value for $e_1^{\rm max}=(-8.3\pm0.14)\times
10^{-5}$. This model is used to correct the observed $e_1$ for stars
and galaxies in our data.

\end{document}